\documentclass{icrc2009}

\usepackage{graphicx}   % for including figures
\usepackage[font=footnotesize]{subfig} % subfig.sty for a double column floating figure using two subfigures
\usepackage{fixltx2e}
\usepackage{url}

\newcommand{\shorttitle}[1]%
{\markboth{Proceedings of the 31\MakeLowercase{$^{st}$} ICRC, {\L}\'{o}d\'{z} 2009}{#1} }
\newcommand{\etal}{\MakeLowercase{\textit{et al. }}} % "et al."

\usepackage{ifthen}
  
\newcommand{\Noff}{\ensuremath{N_\mathrm{off}}}
\newcommand{\Nonn}{\ensuremath{N_\mathrm{on}}}
\newcommand{\BWorCOL}{COL}
\ifthenelse{\equal{\BWorCOL}{COL}}{%if bw:
}{%else:
}

\begin{document}
%\linenumbers

\title{Moon Shadow Observation by IceCube}

\author{\IEEEauthorblockN{D.J.~Boersma\IEEEauthorrefmark{1},
			  L.~Gladstone\IEEEauthorrefmark{2} and
                          A.~Karle\IEEEauthorrefmark{2}\\
                          for the IceCube Collaboration\IEEEauthorrefmark{3}}
                            \\
\IEEEauthorblockA{\IEEEauthorrefmark{1}RWTH Aachen University, Germany}
\IEEEauthorblockA{\IEEEauthorrefmark{2}Department of Physics, University of Wisconsin, Madison, WI 53706, USA}
\IEEEauthorblockA{\IEEEauthorrefmark{3}see special section of these proceedings}
}

% please write the preseter's name and short title (3-4 words maximum)
%    which will appear at the header of the even pages.
\shorttitle{D.J. Boersma \etal IceCube Moon Shadow }
\maketitle

\begin{abstract}
In the absence of an astrophysical standard candle, IceCube can study the
deficit of cosmic rays from the direction of the Moon. The observation of this
``Moon shadow'' in the downgoing muon flux is an experimental verification of
the absolute pointing accuracy and the angular resolution of the detector with
respect to energetic muons passing through. The Moon shadow has been observed
in the 40-string configuration of IceCube. This is the first stage of IceCube
in which a Moon shadow analysis has been successful. Method, results, and some
systematic error studies will be discussed.
\end{abstract}

\begin{IEEEkeywords}
IceCube, Moon shadow, pointing capability
 %please write 3 keywords, to be used to select your subject
 %from all ICRC contributions
\end{IEEEkeywords}
 
%%%%%%%%%%%%%%%%%%%%%%%%%%%%%%%%%%%%%%%%%%%%%%%%%%%%%%%%%%%%%%%
\section{Introduction}
%%%%%%%%%%%%%%%%%%%%%%%%%%%%%%%%%%%%%%%%%%%%%%%%%%%%%%%%%%%%%%%

IceCube is a kilometer-cube scale Cherenkov detector at the geographical South
Pole, designed to search for muons from high energy neutrino interactions.  The
arrival directions and energy information of these muons can be used to search
for point sources of astrophysical neutrinos, one of the primary goals of
IceCube. 

The main component of IceCube is an array of optical sensors deployed in the
glacial ice at depths between 1450~m and 2450~m.  These Digital Optical
Modules (DOMs), each containing a 25~cm diameter photomultiplier tube with
accompanying electronics within a pressure housing, are lowered into the ice
along ``strings.'' There are currently 59 strings deployed of 86 planned; the
data analyzed here were taken in a 40 string configuration, which was in
operation between April 2008 and April 2009.  There are 13 lunar months of data
within that time.  In this analysis we present results from 8 lunar months of 
the 40 string configuration.

 For a muon with energy on the order of a TeV, IceCube can reconstruct
an arrival direction with order $1^{\circ}$ accuracy. For down-going
directions, the vast majority of the detected muons do not originate from
neutrino interactions, but from high energy cosmic ray interactions in the
atmosphere.  These cosmic ray muons are the dominant background in the search
for astrophysical neutrinos. They can also be used to study the performance
of our detector. In particular, we can verify the pointing capability by
studying the shadow of the Moon in cosmic ray muons.

As the Earth travels through the interstellar medium, the Moon blocks some
cosmic rays from reaching the Earth. Thus, when other cosmic rays shower in the
Earth's atmosphere and create muons, there is a relative deficit of  muons 
from the direction of the Moon. IceCube detects these muons, not the primary 
cosmic rays. Since the position and size of the Moon is so well 
known, the resulting deficit can be used for detector calibration. 
The idea of a Moon shadow was first proposed in 1957~\cite{Clark},
and has become an established observation for a number of
astroparticle physics experiments; some examples are given in
references~\cite{moon1,moon2,moon3,moon4}. 
Experiments have used the Moon shadow to calibrate 
detector angular resolution and pointing
accuracy~\cite{tibet}. 
They have also observed the shift of the Moon shadow due to
the Earth's magnetic field~\cite{L3cosmics}. 
The analysis described here is
optimized for a first observation, and does not yet include detailed 
studies such as describing the shape of the observed deficit. These will be 
addressed in future studies.

%%%%%%%%%%%%%%%%%%%%%%%%%%%%%%%%%%%%%%%%%%%%%%%%%%%%%%%%%%%%%%%%%%%%%% 
\section{Method}
%%%%%%%%%%%%%%%%%%%%%%%%%%%%%%%%%%%%%%%%%%%%%%%%%%%%%%%%%%%%%%%%%

\subsection {Data and online event selection}

Data transfer from the South Pole is limited by the bandwidth of two 
satellites; thus, not all downgoing muon events can be immediately transmitted.
This analysis uses a dedicated online event selection, choosing
events with a minimum quality and a reconstructed direction within a
window of acceptance around the direction of the Moon.  The
reconstruction used for the online event selection is a single
(i.e., not iterated) log-likelihood fit.
%Both fits use a simple analytic model of the scattering of light in ice.

The online event selection is defined as follows, where
 $\delta$ denotes the Moon declination:
\begin{itemize}
\item The Moon must be at least $15^{\circ}$ above the horizon. 
\item At least 12 DOMs must register each event.
\item At least 3 strings must contain hit DOMs. 
\item The reconstructed direction must be within 10$^{\circ}$ of the Moon 
in declination.
\item The reconstructed direction must be within 40$^{\circ}/\cos(\delta)$ of 
the Moon in right ascension;
the $\cos(\delta)$ factor corrects for projection effects.
\end{itemize}
These events are then sent via satellite to the northern hemisphere for 
further processing, including running the higher-quality 32-iteration 
log-likelihood reconstruction used in further analysis.

\begin{figure}[!t]
 \centering
 \includegraphics[width=3.0in]{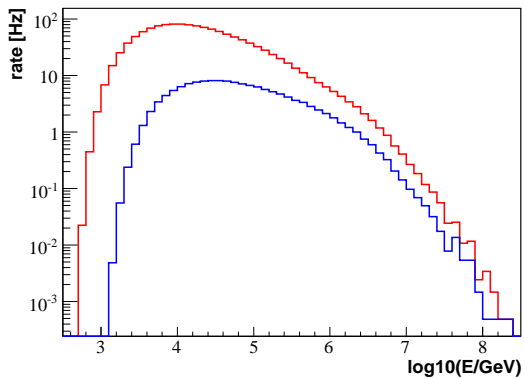}
 \caption{The energy spectrum of (simulated) CR primaries of muons (or muon
bundles) triggering IceCube. Red: all events;
blue: primaries with $\delta>-30^{\circ}$.}
 \label{EcrM}
\end{figure}

The Moon reached a maximum altitude of $27^{\circ}$ above the horizon
($\delta=-27^{\circ}$) in 2008, when viewed from the IceCube detector.
The trigger rate from cosmic ray
muons is more than 1.2~kHz in the 40 string configuration, but
most of those muons travel nearly vertically, and thus they cannot 
have come from directions near the Moon.
Only $\sim11\%$ of all muons that trigger the detector 
come from angles less than $30^{\circ}$ above the horizon.  
Furthermore, muons which are closer to horizontal (and thus closer to the 
Moon) must travel farther before reaching the detector. 
They need a minimum energy to reach this far (see Fig.~\ref{EcrM}):
 the cosmic ray primaries which produce them must have energies of at 
least 2~TeV.

% see \section{Examples of \LaTeX\  instructions} and \subsection{Figures}
\begin{figure}[!t]
 \centering
 \includegraphics[width=3.0in]{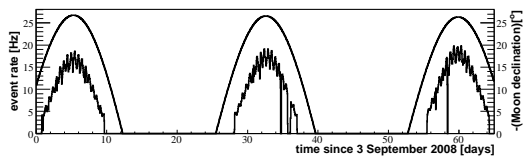}
 \caption{The rate of events passing the Moon filter (in Hz, lower curve) 
averaged hourly, together with the position of the Moon above the horizon
 at the South Pole (in degrees, upper curve), plotted versus time over 3 
typical months.}
 \label{livetime}
\end{figure}

Three typical months of data are shown in Fig.~\ref{livetime}, along with the
position of the Moon above the horizon.  The dominant shape is from the strong
increase in muon flux with increasing angle above the horizon: as the Moon
rises, so do the event rates near the Moon. This can be seen clearly in the
correlation between the two sets of curves.  There is a secondary effect from the
layout of the 40 strings. One dimension of the detector layout has the full
width (approximately 1km) of the completed detector, while the other is only
about half as long. When the Moon is aligned with the short axis, fewer events
pass the filter requirements. This causes the 12~hour modulation in the rate.

%%%%%%%%%%%%%%%%%%%%%%%%%%%%%%%%%%%%%%%%%%%%%%%
\subsection{Optimization of offline event selection and search bin size}

A simulated data sample of $10^5$ downgoing muon events was
generated using CORSIKA~\cite{CORSIKA}.

% see \section{Examples of \LaTeX\  instructions} and \subsection{Figures}
\begin{figure}[!t]
 \centering
 \includegraphics[width=3.0in]{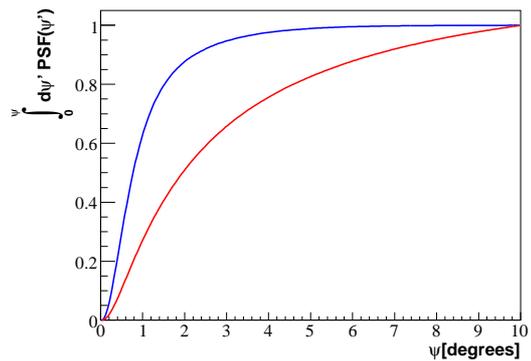}
 \caption{The x-axis shows the angular difference $\psi$ between the true and
 reconstructed track. The y-axis shows the fraction of events with this or
 lower angular error. The blue curve shows the event sample after offline
 event selection, and the red curve shows the event sample after online 
event selection.}
 \label{psf}
\end{figure}

% see \section{Examples of \LaTeX\  instructions} and \subsection{Figures}
\begin{figure*}[th]
 \centering
 \includegraphics[width=6in]{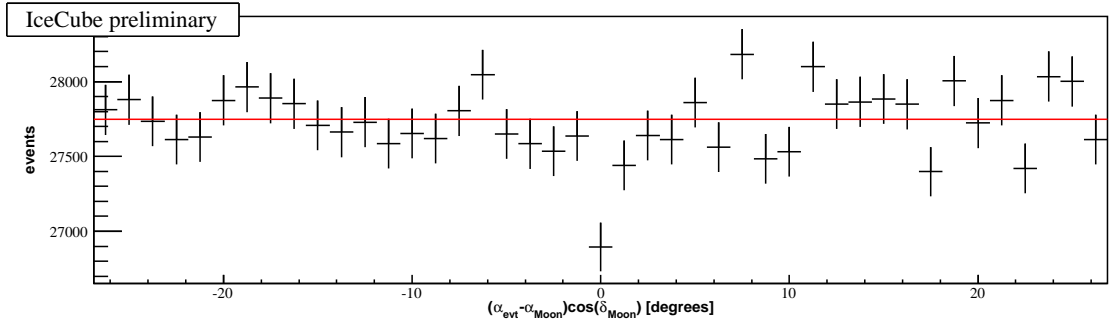}
 \caption{Number of events per $1.25^{\circ}$ square bin, relative to the
 position of the Moon. The declination of the reconstructed track is within
 $0.625^{\circ}$ bin from the declination of the Moon.  The average of all bins
 except the Moon bin is shown as a red line to guide the eye.}
 \label{rawdata}
\end{figure*}

\begin{figure*}[th]
 \centering
 \includegraphics[width=6in]{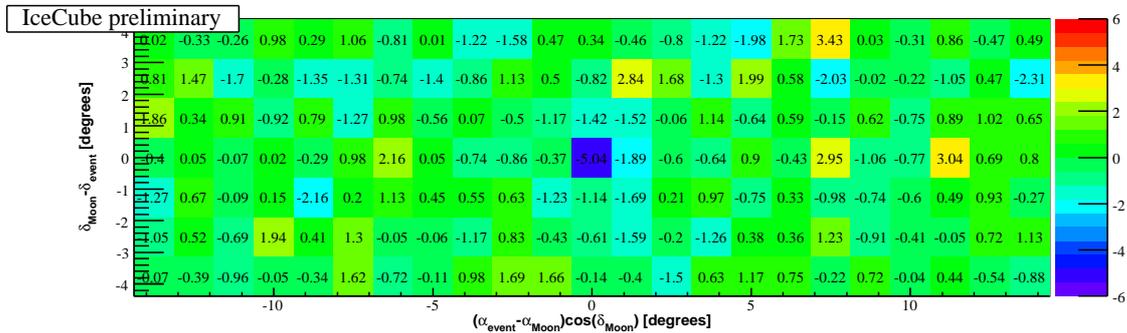}
 \caption{The significance of deviations in a region 
centered on the Moon.}
 \label{signif}
\end{figure*}

\begin{figure}[!t]
 \centering
 \includegraphics[width=3.0in]{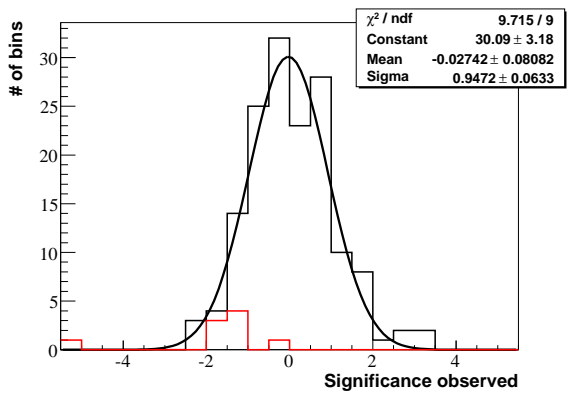}
 \caption{Each of the deviations shown in Fig.~\ref{signif} is plotted here.
 The deviations of the central 9 bins are shown in red.  The
 surrounding bins are shown with a black line histogram, and fit with a
 Gaussian curve.}
 \label{gaus}
\end{figure}

A set of cuts was developed using the following estimated relation between 
the significance $S$, the
efficiency $\eta$ of events passing the cut, 
and the resulting median angular resolution $\Psi_\mathrm{med}$ of the sample:

\begin{equation}
S(\mathrm{cuts}) \propto \frac{\sqrt{\eta(\mathrm{cuts})}}{\Psi_\mathrm{med}(\mathrm{cuts})}
\end{equation}
Since the  deficit is based on high statistics of events in the search bin,
this function provides a good estimator for optimizing the significance.  

The following cuts were chosen: 
\begin{itemize} 
\item At least 6 DOMs are hit with light that hasn't been scattered in the ice,
allowing a -15 nsec to +75 nsec window from some minimal scattering.
%\item At least 6 DOMs are hit directly, that is,
%the arrival time of the hits was within a -15~nsec to +75~nsec window
%from the calculated Cherenkov photon arrival time (allowing some minimal scattering).
\item  Projected onto the reconstructed track, two of those hits
at least 400 meters apart.
\item  The $1\sigma$ estimated error ellipse on the reconstructed direction
 has a mean radius less than $1.3^{\circ}$.
\end{itemize} 
%%% Add me back in !! But don't leave a space, it makes it start a new pgh.
%In simulation, these cuts retain xx\% of events, with a 
%median angular resolution of xx. 
The cumulative point spread function 
of the sample after the above quality cuts is shown as 
the blue line in Fig.~\ref{psf}. 

The size $\Psi_\mathrm{search}$ of the search bin is optimized for a
maximally significant observation using a similar $\sqrt{N}$-error based 
argument and the resulting relation, which follows. Using the cumulative 
point spread function of the sample after quality cuts, we have:

\begin{equation}
S(\Psi_\mathrm{search}) \propto 
\frac{ \int_0^{\Psi_\mathrm{search}}PSF(\psi^\prime)d\psi^\prime}{\Psi_\mathrm{search}}
\end{equation}
Maximizing this significance estimator gives an optimal search bin radius of
$0.7^{\circ}$. This analysis uses square bins with an area equal to that of the
optimized round bin, with side length $1.25^{\circ}$.

%%%%%%%%%%%%%%%%%%%%%%%%%%%%%%%%%%%%%%%%%%%%%%%%%%%%%%%%%
\subsection{Calculating significance}
\label{section:significance}

To show that the data are stable in right ascension $\alpha$, we show, in
Fig.~\ref{rawdata}, the number of events in the central declination band.  The
errors shown are $\sqrt{N}$. The average of all bins excluding the Moon bin is
27747, which is plotted as a line to guide the eye. The Moon bin has 852 events
below this simple null estimate. This represents a $5.2\sigma$ deficit using
$\sqrt{N}$ errors. 

Although this shows that the data are stable, this error system is 
vulnerable to variations in small data samples. Although we don't
see such variations here, we considered it prudent to consider 
an error system which takes into account the size of the background sample. 

% one figure used to live here-- moved for better layout

We used a standard formula from Li and Ma~\cite{LiNMa} for calculating the
significance of a point source:

\begin{equation}
 S=\frac{\Nonn-\alpha \Noff}{\sqrt{\alpha(\Nonn+\Noff)}}.
\end{equation}
where \Nonn\ is the number of events in the signal sample, \Noff\ is the number
of events in the off-source region, and  $\alpha$ is the ratio between
observing times on- to off-source. We take $\alpha$ instead as the ratio of on-
to off-source areas observed, since the times are equal.

%%Two figures were orignally here, but I've moved them around so they 
%% show up earlier.

The above significance formula is applied to the Moon data sample in the
following way. The data are first plotted in the standard Moon-centered
equatorial coordinates, correcting for projection effects with a factor of
$\cos(\delta)$. The plot is binned using the $1.25^{\circ} \times 1.25^{\circ}$
bin size optimized in the simulation study. Each bin successively is considered
as an on-source region. There is a very strong declination dependence in the
downgoing muon flux, so variations of the order of the Moon deficit are
only detectable in right ascension. Thus, off-source regions are selected
within the same zenith band as the on-source region. Twenty off-source bins are
used for each calculation: ten to the right and ten to the left of the
on-source region, starting at the third bin out from the on-source bin (i.e.,
skipping two bins in between).

%%%%%%%%%%%%%%%%%%%%%%%%%%%%%%%%%%%%%%%%%%%%%%%%%%%%%%%%%%%%%%%%%%%%%%%
\section{Results}
%%%%%%%%%%%%%%%%%%%%%%%%%%%%%%%%%%%%%%%%%%%%%%%%%%%%%%%%%%%%%%%%%%%%%%%%

For a region of 7 bins or $8.75^{\circ}$ in declination $\delta$ and 23 bins or
$28.75^{\circ}$ in right ascension $\alpha$ around the Moon,
the significance of the deviation of the count rate in each bin with respect to
its off-source region was calculated,
as described in section~\ref{section:significance}.
The result is plotted in 
Fig.~\ref{signif}.  The Moon can be seen as the $5.0\sigma$ deficit in the
central bin, at $(0,0)$.

To test the hypothesis that the fluctuations in the background away from the
Moon are distributed randomly around 0, we plot them in Fig.~\ref{gaus}. The
central 9 bins, including the Moon bin, are not included in the Gaussian fit,
but are plotted as the lower, shaded histogram.
The width of the Gaussian fit is consistent with 1; therefore, the background
is consistent with random fluctuations.

%%%%%%%%%%%%%%%%%%%%%%%%%%%%%%%%%%%%%%%%%%%%%%%%%%%%%%%%%%%%%%%%%%%%
\section{Conclusions and Future Plans}
%%%%%%%%%%%%%%%%%%%%%%%%%%%%%%%%%%%%%%%%%%%%%%%%%%%%%%%%%%%%%%%%%%%%%%%%
IceCube has observed the shadow of the Moon as a $5.0\sigma$ deviation from
event counts in nearby regions, using data from 8 of the total 13 lunar months
in the data taking period with the 40-string detector setup. From this, we can
conclude that IceCube has no systematic pointing error larger than the search
bin, $1.25^{\circ}$. 

In the future, this analysis will be extended in many ways. First, we will 
include all data from the  40 string detector configuration.  
We hope to repeat this analysis using unbinned likelihood methods, and 
to describe the size, shape, and any offset of the Moon Shadow.  We will then
use the results of these studies to comment in more detail on the 
angular resolution of various reconstruction algorithms within IceCube. This 
analysis is one of the only end-to-end checks of IceCube systematics based 
only on experimental data. 

LG acknowledges the support of a 
National Defense Science and Engineering Graduate Fellowship 
from the American Society for Engineering Education.

\end{document}